\documentclass[
10pt,
showpacs,preprintnumbers,footinbib,
amsmath,amssymb,
aps,
prl,twocolumn,groupedaddress,superscriptaddress,
showkeys
]{revtex4-1}
\usepackage{graphicx}
\usepackage{dcolumn}
\usepackage{bm}
\usepackage[colorlinks=true,urlcolor=blue,citecolor=blue]{hyperref}
\usepackage{color}

\newcommand{\nmax}{N_{\rm max}}
\newcommand{\hb}{\hbar\Omega}
\newcommand{\ncsm}{No Core Shell Model }

\begin{document}

\title{{\em Ab-initio} calculation of the photonuclear cross section of $^{10}$B}

\author{M.K.G. Kruse}
\email{kruse9@llnl.gov}
   \affiliation{Lawrence Livermore National Laboratory, P.O. Box 808, L-414, Livermore, California 94551, USA}

\author{W.E. Ormand}
\affiliation{Lawrence Livermore National Laboratory, P.O. Box 808, L-414, Livermore, California 94551, USA}

\author{C.W. Johnson}
\affiliation{San Diego State University, 5500 Campanile Drive, San Diego, California 92182, USA}

\date{\today}

\begin{abstract}
\noindent We present for the first-time the photonuclear cross section of $^{10}$B calculated within the {\em ab-initio} \ncsm framework. Realistic two-nucleon (NN) chiral forces up to next-to-next-to-next-order (N3LO), which have been softened by the similarity renormalization group method (SRG) to $\lambda=2.02$ fm$^{-1}$, were utilized. The electric-dipole response function is calculated using the Lanczos method. The effects of the continuum were accounted for by including neutron escape widths derived from R-matrix theory. The calculated cross section agrees well with experimental data in terms of structure as well as in absolute peak height, $\sigma_{\rm max}=4.85~{\rm mb}$ at photon energy $\omega=23.61~{\rm MeV}$, and integrated cross section $85.36\, {\rm MeV \cdotp mb}$. We test the Brink hypothesis by calculating the electric-dipole response for the first five positive-parity states in $^{10}$B and verify that dipole excitations built upon the ground- and excited states have similar characteristics.

\end{abstract}

\pacs{}

\maketitle


Electric-dipole transitions are an important excitation mode characterizing many facets of nuclear structure. Of particular interest is their strongly collective nature, which is manifested in what is known as the giant-dipole resonance (GDR). The GDR is ubiquitous in nuclei, and since its initial observation~\cite{Baldwin47}, much experimental and theoretical effort has been devoted to understanding its properties. The centroid of the GDR generally scales as the inverse of the nuclear radius, and experimentally is found to be $\sim 79 A^{-1/3}$~MeV, while the width is of the order 5~MeV~\cite{Bohr69-vol2}.  Early on, phenomenological models were proposed by Goldhaber and Teller \cite{Goldhaber48} and Steinwedel and Jensen \cite{Steinwedel50} based on proton-neutron fluids that were able to describe the energy of the resonance. The width, on the other hand, was postulated to be due to the GDR damping into other nuclear modes of motion~\cite{Speth81,Bertsch83}.  A further, intriguing property is that a collective dipole mode exists on each state of the nuclear system, as hypothesized by Brink~\cite{Brink55}. The resonant part of the photonuclear cross section has been calculated with semi-realistic interactions for $^4$He \cite{Efros97,Stetcu07}, $^6$He and $^{6}$Li \cite{Bacca02,Bacca04} and $^{7}$Li \cite{Bacca04-02}. More recent calculations for $^4$He utilizing modern realistic two- and three-body interactions have also been performed \cite{Gazit06,Quaglioni07}. These calculations have either used the framework of the hyperspherical harmonics (HH) expansion \cite{Barnea01,Fenin72} or the \ncsm (NCSM) \cite{Nav00,NavRev09,Barrett13}. Recently the Lorentz integral method \cite{Efros94,Efros07} was used in conjunction with coupled-cluster calculations to calculate the $^{16}$O giant dipole resonance \cite{Bacca13}.

In this letter, we report on a theoretical study of the properties of the GDR for $^{10}$B within the framework of the ab initio No Core Shell Model (NCSM)~\cite{Nav00,NavRev09,Barrett13}. We study the convergence properties of the GDR as a function of the model space size and present the photo-nuclear absorption cross section for $^{10}$B. We demonstrate the influence of more complex modes on the damping of the GDR as well as the influence of the neutron escape width on the dipole response. Finally, we test the Brink hypothesis by calculating the dipole response on positive parity excited states in $^{10}$B and find a robust GDR built on each of these states exhibiting remarkably similar properties.

The NCSM is a bound-state technique appropriate for light nuclei that uses as input realistic two- and three-body nuclear interactions. The NCSM determines the eigenenergies and wave functions of the nucleus by expressing the translationally invariant Hamiltonian in terms of antisymmetric combinations of single-particle harmonic oscillator (HO) states of frequency $\Omega$. The size of the Slater determinant basis is determined by the total HO quanta, $\nmax$, available in the system above the lowest configuration. Realistic interactions that make the link between structure and quantum chromodynamics explicit are derived using the effective field theory (EFT) for nuclear forces~\cite{Epelbaum09,Machleidt11}. In this work, we include nucleon-nucleon (NN) terms up to next-to-next-to-next-leading order (N3LO)~\cite{EM500}. To enhance convergence, an effective interaction was employed using the similarity renormalization group procedure 
(SRG)~\cite{Bogner07,Jurgenson09,Bogner10,Roth11} with a momentum-decoupling value of $\lambda=2.02$~fm$^{-1}$. At this $\lambda$ value, the binding energies of $p$-shell nuclei are reproduced as though the calculation was performed with both the N3LO NN and N2LO NNN interactions~\cite{Jurgenson11,Jurgenson13,Baroni13-PRC}. In order to isolate effects of the strong interaction, we use an isospin-symmetric (isoscalar) interaction, and ignore the Coulomb interaction. All calculations were performed with $\hb=20$~MeV.

The dipole response function $S(\omega)$ on an initial state with angular momentum $J$ and energy $E$ is given by
\begin{eqnarray}
	S(\omega) &=& \frac{1}{2J+1} \sum_{f,M} |\langle J_f M_f | \hat{D}_z | J M \rangle |^2 \delta(E_f - E - \omega) \nonumber \\
				&=& \sum_f \frac{B(E1; J\rightarrow J_f)}{3} \delta(E_f - E - \omega),
	\label{eq:reduced-strength}
\end{eqnarray}
where we assume the photon polarization is in the $z$-direction, the sum is taken over all initial orientations $M$ and final states $f$, and $\omega$ is the photon energy. $D_\mu$ is the translationally-invariant dipole operator
\begin{equation}
	\hat{D}_\mu= \sum_{i=1}^A {r}_i Y_{1\mu}(\Omega_i) \left(1/2-Z/A+\hat{t}_{z,i}\right),
	\label{eq:dipole-operator}
\end{equation}
with the $Z$, ${r}_i$ and $\Omega_i$ denoting the proton number, single-particle radial and angular coordinates, respectively, and $\hat{t}_{z,i}$ is the z-component of the isospin operator whose eigenvalues are $\pm1/2$ when applied to a proton/neutron, respectively~\cite{Brussaard}. Finally, $B(E1; J \rightarrow J_f)$ is the reduced transition probability. 
 

The response function, $S(\omega)$, was computed using the Lanczos algorithm as described in Refs.~\cite{Caurier90,Caurier95,Caurier99,Caurier05,Whitehead80,Bloom84}. First, $D_z$ is applied to an eigenstate and angular momentum $J_f$ is then projected on this vector. This new vector is then used as an initial pivot for the Lanczos procedure, and $n$ Lanczos iterations are performed giving $n$ eigenvalues with energy $E_f$. The reduced transition probability is then extracted from the overlap between the initial pivot and the Lanczos eigenvectors, and Eq.~(\ref{eq:reduced-strength}) is evaluated. We note that of these $n$ Lanczos eigenvalues, only a subset of them is fully converged (generally just the extreme values), but with Eq.~(\ref{eq:reduced-strength}), these $n$ Lanczos eigenvalues reconstruct the first $2n-1$ moments of $S(\omega)$. $S(\omega)$ was computed with $n$ up to 300.
All of the many-body calculations, including the Lanczos calculation of the strength function
distribution, were carried out using the BIGSTICK configuration-interaction code~\cite{BIGSTICK}. 

\begin{figure}[h]
\begin{center}
	\includegraphics{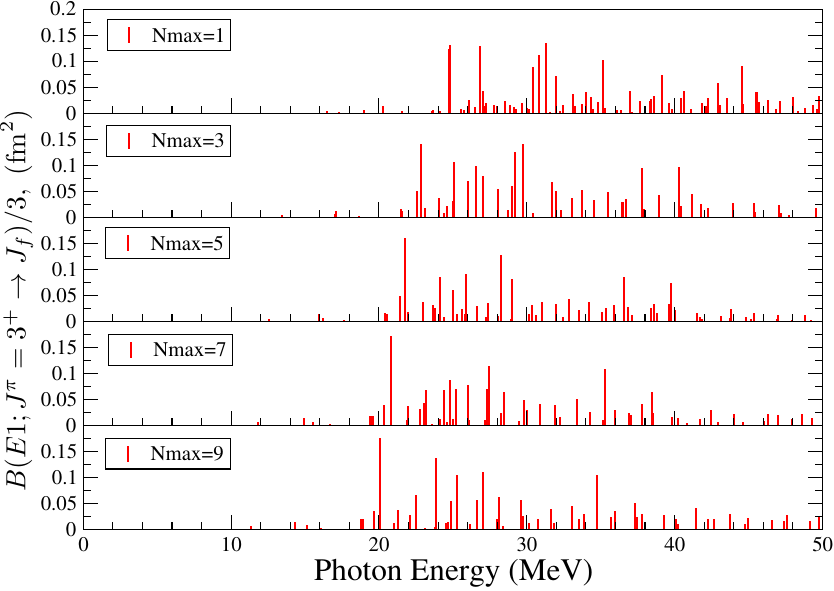}
	\caption{(color online) The components of the strength function for the $^{10}$B ground-state ($J^\pi = 3^+$) as a function of increasing basis size $\nmax$ at $\hb=20~{\rm MeV}$. The $\nmax=5-9$ strength functions were calculated with 300 iterations. }
	\label{fig:strength-function-nmax}
\end{center}
\end{figure}

The $E1$ response function for the ground-state of $^{10}$B is shown in Fig. \ref{fig:strength-function-nmax} as a function of increasing basis size $\nmax$. Overall, the GDR excitation spectrum shifts to lower energy with increasing $\nmax$, reflecting a somewhat slower convergence for these negative-parity excited states relative to the $^{10}$B ground state. 


Fig.~\ref{fig:strength-function-nmax} illustrates one of the principal broadening mechanisms for the GDR. In particular, as $\nmax$ increases, the simple collective 1-particle, 1-hole (1p-1h) nature of the GDR couples to more complex states, and the strength ``spreads out'' across a wider range of states. In particular, the GDR is ``damped'' via mixing with these more complex low-lying 3p-3h (and higher)  states. 

A second broadening mechanism must be addressed in order to compare with experimental data. This arises from the fact that most of the states excited by the dipole operator lie above the threshold for particle emission; in particular neutrons at $S_n = 8.436$~MeV~\cite{Tilley04}. Thus, the dominant decay mode of these states is neutron emission, which occurs on a short time scale. We can estimate the width of these excited states with the neutron-escape width, $\Gamma_n$, from the R-matrix theory
\begin{equation}
	\Gamma_n(E_x)=2\gamma_{sp}^2 \sum_l P_l \theta_{l}^2,
	\label{eq:neutron-escape-width}
\end{equation}
where $E_x$ is the excitation energy, $\gamma_{sp}^2=({\hbar c})^2/{\mu R^2}$ is the single-particle width, $\mu=(({A-1})/{A})m_N$ is the reduced mass of the $A$-nucleon system in terms of the nucleon mass $m_N$, and the radius is $R=1.2[(A-1)^\frac{1}{3}+1]$. The penetration factor $P_l=k_nR V_l$ depends on the wave-vector $k_n=\sqrt{2\mu E_n}$ ($E_n$ is the neutron kinetic energy) and the penetrability, $V_l$, of the outgoing neutron with orbital angular momentum $l$. Since $^{10}$B is a p-shell nucleus and the GDR is dominated by 1p-1h excitations from the $0p$ shell into the $0d-1s$ shell, neutron emission of these states will be dominated   $l=0$ and $l=2$ partial waves. For $s$-waves, $V_0 =1$, while for $d$-waves, $V_2 = {(kR)^4}/{(9+3(kR)^2+(kR)^4)}$~\cite{atlas06}.

\begin{figure}[h]
\begin{center}
	\includegraphics{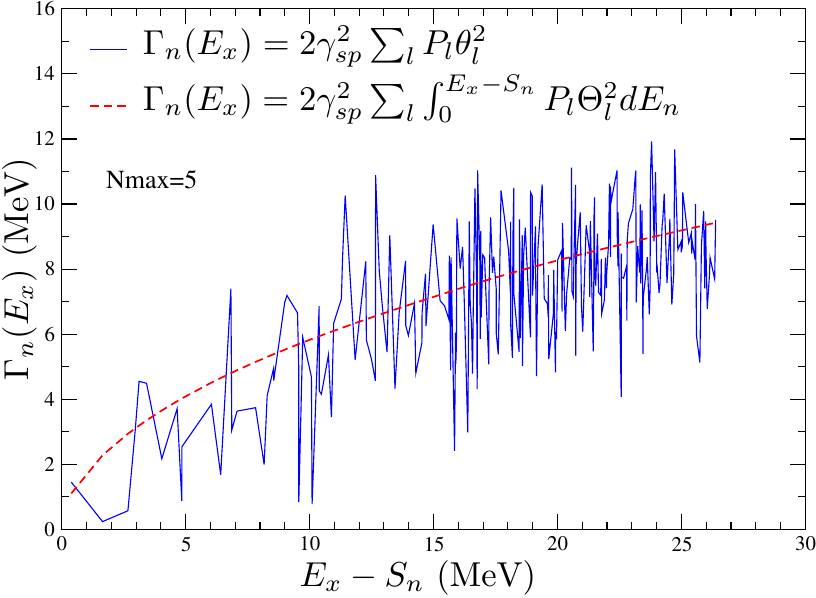}
	\caption{(color online) The widths of $^{10}$B negative parity states as a function of the excitation energy ($E_x$) above the neutron separation energy ($S_n$) are shown by the blue (solid) line when the spectroscopic factors are taken into account as in Eq. (\ref{eq:neutron-escape-width}) whereas the red (dashed) line shows the result of using a model of the spectroscopic factors as in Eq. (\ref{eq:neutron-escape-width-model}). The calculation is performed at $\nmax=5$. }
	\label{fig:model-gamma-vs-real-gamma}
\end{center}
\end{figure}

The relative strength for each partial wave in Eq. (\ref{eq:neutron-escape-width}) is given by the spectroscopic factor,
\begin{equation}
	\theta_l=\left[ \langle ^9B; J_d T_d| \times \langle n; (nlj;1/2)|\right]^{J_fT_f} |^{10}B; J_fT_f \rangle,
\end{equation}
which is a measure of the overlap of the neutron$+^9$B with the neutron-unbound $^{10}$B (the neutron is in the HO orbital state $(nlj)$ with isospin $t=1/2$). The magnitude of the width is strongly governed by the outgoing kinetic energy ($\sim \sqrt{E_n}$), which in turn is governed by the distribution of the spectroscopic strength to states in 
$^9$B ($E_n = E_x -S_n-E_d$). Indeed, in the simplest approximation, where the entire spectroscopic strength is collected in the ground state of $^9$B, $\Gamma_n$ increases dramatically with excitation energy, and is unphysical. On the other hand, due to the large dimensions present it is also not practical to calculate the spectroscopic strength in the largest model spaces and for highly excited states. Instead, we build a model for the spectroscopic strength,  $\Theta^2_l (E_n+S_n)$, based on fairly complete calculations for $\nmax=3$ and $5$.  The spectroscopic strength is modeled with a normalized Gaussian that reproduces the total strength as well as the first and second moments of the calculated $\nmax=3$ and $5$ results, which is feasible up to $E_x \sim 30$~MeV. The neutron-escape width is then estimated by
\begin{equation}
	\Gamma_n(E_x)=2\gamma_{sp}^2 \sum_l \int_0^{E_x-S_n} P_l \Theta_{l}^2(E_n+S_n) dE_n.
	\label{eq:neutron-escape-width-model}
\end{equation}
In Fig. \ref{fig:model-gamma-vs-real-gamma}, we compare the actual widths calculated by Eq. (\ref{eq:neutron-escape-width}) with those of the model of spectroscopic factors from Eq. (\ref{eq:neutron-escape-width-model}). We find that $\Gamma_n$  increases with excitation energy, but much less dramatically than if the decay occurred to the ground state. Further, the width appears to saturate to a value of 10-12 MeV at $E_x\approx 40-50$~MeV. This is largely due to the fact that the distribution of spectroscopic strength occurs to states with one less oscillator quanta than the initial state, and thus tends to cap the outgoing kinetic energy at roughly $\hbar\Omega$. For what follows, we compute the neutron escape width using the model for the spectroscopic strength constrained by the $\nmax=5$ calculation. 

\begin{figure}[h]
\begin{center}
	\includegraphics{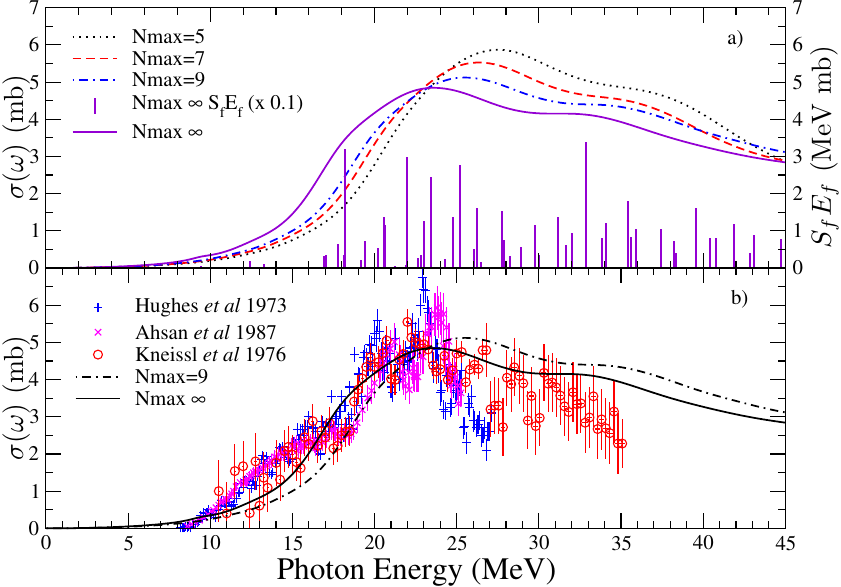}
	\caption{(color online) a) The photonuclear cross section as a function of $\nmax=5-9, \infty$. Note that the main peak slowly shifts down in photon energy ($\omega$) as well as decreases in height. The energy-weighted strength, $S_f E_f$, is shown for $\nmax=\infty$ with vertical lines. b) The calculated cross section at $\nmax=\infty$ (solid) is compared to three sets of experimental photoneutron cross section data (symbols). Our absolute peak height of $\sigma_{\rm max}=4.85~{\rm mb}$ at $\omega=23.61~{\rm MeV}$ compares well with the Kneissl data.} 
	\label{fig:cross-section-nmax9}
\end{center}
\end{figure}

The photo-absorption cross section in the unretarded dipole-approximation is related to the strength function by $\sigma(\omega)=4\pi^2 \alpha \omega \left( S(\omega)-S(-\omega) \right)$, where $\alpha=\frac{e^2}{\hbar c}$ is the fine-structure constant \cite{Bohr69-vol2}. For states above the neutron separation energy, the discrete lines in $S(\omega)$ were replaced with a normalized Breit-Wigner function centered at the eigen-energy $E_f$ with width $\Gamma_n(E_f)$. The photo-absorption cross section is then 
\begin{equation}
	\sigma(\omega)=4 \pi^2 \alpha \sum_f S_f E_f \frac{\omega^2 \Gamma_n(E_f)}{(\omega^2-E_f^2)^2+(\omega \Gamma_n(E_f))^2},
	\label{eq:CS-used}
\end{equation}
with $\Gamma_n(E_f)$ given by Eq.~(\ref{eq:neutron-escape-width-model}) as shown in Fig.~\ref{fig:model-gamma-vs-real-gamma} and $S_f={B(E1; J\rightarrow J_f)}/{3}$.
The photo-absorption cross section calculated for $^{10}$B is shown in Fig.~\ref{fig:cross-section-nmax9} a), for $\nmax=5-9$ as well as an extrapolation to $\nmax=\infty$ . The vertical lines show the discrete spectrum at $\nmax=\infty$, while the continuous lines show the cross section with the neutron-escape width folded into the spectrum. The $\nmax=\infty$ result was obtained with an exponential extrapolation~\cite{Nav04}, where the absolute energies of the ground state and the first large peak in the dipole-response function shown in Fig. \ref{fig:strength-function-nmax} (see the peak at 20~MeV for $\nmax=9$) for $\nmax=3-9$ were fit to the function $E(\nmax)=A\exp(-B\nmax)+C$. The parameter $C$ is then the energy at $\nmax=\infty$. This procedure leads to a final excitation energy for the fitted dipole state of 18.209~MeV, which is 1.881~MeV lower than the $\nmax=9$ result. To estimate the $\nmax=\infty$ strength function, we assume this shift to be uniform across the entire spectrum, and shift the $\nmax=9$ dipole response lower by 1.881~MeV.
Overall, the cross section is seen to have a main peak at about 23 MeV for $\nmax=\infty$, followed by a slightly smaller hump near 33 MeV. We also note that the main peak in the spectrum tends to shifts down in photon energy ($\omega$) as well as decreasing in height as the model space increases.  


In Fig. \ref{fig:cross-section-nmax9} b), we compare the $\nmax=9$ and $\nmax=\infty$ calculation with  experimental data.
Hughes {\em et al} \cite{Hughes73} and Ahsan {\em et al} \cite{Ahsan87} determine the cross section from bremsstrahlung sources, whereas Kneissl {\em et al} \cite{Kneissl76} used a quasi-monoenergetic photon beam from positron annihilation in flight. The bremsstrahlung data is in good agreement with each other whereas the Kneissl data is about 10\% lower at the peak and has a slower fall-off in the tail of the cross section. In Ref. \cite{Kneissl76} it is argued that cross sections determined from bremsstrahlung sources have to be extracted from yield curves by an unfolding procedure, which may introduce spurious structure in the cross section. Furthermore, comparison between absolute cross sections for $^6$Li and $^7$Li measured at Livermore \cite{Berman74,Berman73} using monoenergetic photons and those by Hayward \cite{Hayward65} with bremsstrahlung sources are also about 10\% lower. The peak in our calculated cross section is slightly higher in photon energy than experimental data by about 1.5 MeV. The absolute peak height of $\sigma_{\rm max}=4.85~{\rm mb}$ at $\omega=23.61~{\rm MeV}$ compares well with the Kneissl data and shows a similar pattern in the tail of the cross section. The $\nmax=\infty$ integrated cross section, $\int_0^{35} \sigma(\omega)d\omega=85.36\, {\rm MeV \cdotp mb}$ compares well with the experimental value of $83.1\pm1.2 \, {\rm MeV \cdotp mb}$ \cite{Kneissl76}. Another evident feature is that both the experimental data and the theory calculation bear out a two-hump structure in the cross section implying that $^{10}$B may be deformed. At $\nmax=9$, our calculated quadrupole moment for the $J^\pi=3^+$ state is $Q=6.22\, {\rm fm}^2$. Using a hydrodynamical model of the nucleus \cite{Neyens03} one can estimate the deformation parameter $\beta_{2,{\rm NCSM}} \approx \sqrt{5\pi}Q/(3 Z R^2) = 0.37$, where we have taken our calculated radius as $R=2.11\,{\rm fm}$ \footnote{The calculated rms radius is for point-nucleons and thus should not be directly compared to experimental data. Since the quadrupole moment is similarly calculated we expect that finite-size effects will be canceled in the ratio $Q/R$.}. A similar calculation using the experimental data of $Q=8.477\, {\rm fm}^2$ \cite{Tilley04} and rms charge radius $R=2.58\, {\rm fm}$ \cite{Cichocki95} leads to $\beta_{2,{\rm exp}}=0.34$.  Varying the fit parameters of the spectroscopic factor model changes the peak-height of the cross section by about 10\% but does not affect the threshold or tail region of the calculated cross section.

\begin{figure}[h]
\begin{center}
	\includegraphics{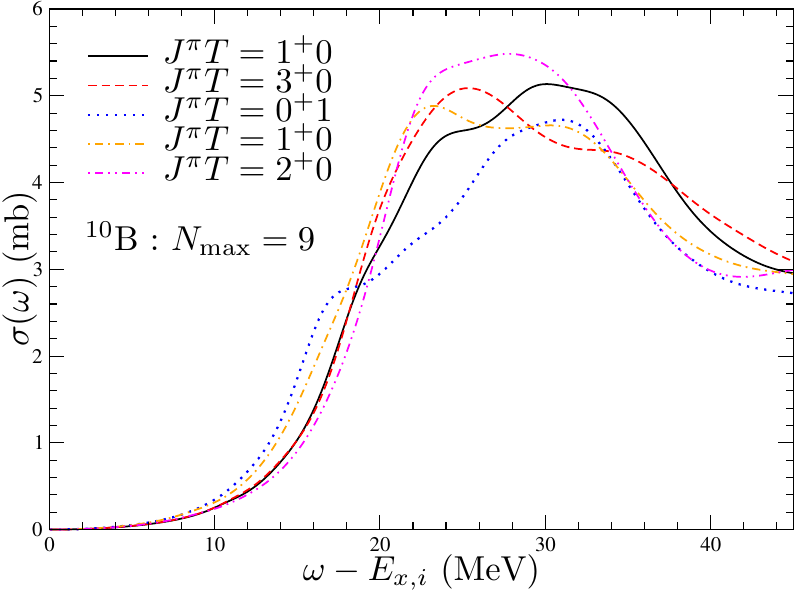}
	\caption{(color online) The dipole response of the first five states in $^{10}$B is shown. The dipole response of the excited states has been shifted down in photon energy by their respective excitation energy $E_{x,i}$. } 
	\label{fig:Brink}
\end{center}
\end{figure}

Lastly, we address the Brink hypothesis, which postulates that if the GDR is observed for the ground-state, then all excited states will similarly support a dipole resonance~\cite{Brink55}. In Fig. \ref{fig:Brink}, we show results of the GDR photo-absorption cross  calculated on the five lowest states in the NCSM spectrum for $^{10}$B relative to the energy of the initial state. Our calculations give remarkably similar cross sections for each of the states, with a strong collective mode at approximately the same excitation energy and width. We thus report the first fully microscopic calculation, based on realistic nucleon-nucleon interactions, that demonstrate the universal feature of the giant-dipole resonance in nuclei. 

In conclusion, we have presented the first NCSM calculations of the dipole response function of $^{10}$B using realistic two-body chiral N3LO interactions that have been softened by SRG to $\lambda=2.02 \, {\rm fm^{-1}}$. To determine the photonuclear cross section we assigned finite widths to the continuum-discretized spectrum by using the physics of neutron-escape widths from R-matrix theory. A model was built to mimic the behavior of the spectroscopic factors as a function of the excitation energy of the $^{10}$B particle-unbound states. The calculated cross section agrees well with the experimental data in terms of structure as well as in absolute peak height $\sigma_{\rm max}(23.61)=4.85~{\rm mb}$ and integrated cross section $85.36\, {\rm MeV \cdotp mb}$. Our calculations support the Brink hypothesis and illustrate that excited states may have a different GDR shape than that of the ground-state. The methodology presented here represents the first step in creating a NCSM framework for calculating electroweak observables for light nuclei in general.

M.K.G.K would like to thank S.A. Coon, J. Escher and W. Haxton for stimulating discussions regarding this work. We thank A. Tonchev for discussing with us various experimental data regarding giant dipole resonances. Computing support for this work came from the Lawrence Livermore National Laboratory (LLNL) institutional Computing Grand Challenge program. It was prepared by LLNL under Contract No. DE-AC52-07NA27344. This material is based upon work supported by the U.S. Department of Energy, Office of Science, Office of Nuclear Physics, under Work Proposal No. SCW0498 
and Award Number  DE-FG02-96ER40985.    

\bibliography{B10-prl}


\end{document}